# Trends in valence band electronic structure of mixed uranium oxides

Kristina O. Kvashnina*[a,b], Piotr M. Kowalski *[c,d], Sergei M. Butorin [e], Gregory Leinders[f], Janne Pakarinen[f], Rene Bes[g], Haijian Li[c,d], Marc Verwerft [f]

[a.] Rossendorf Beamline at ESRF – The European Synchrotron, CS40220, 38043 Grenoble Cedex 9, France
[b.] Helmholtz Zentrum Dresden-Rossendorf (HZDR), Institute of Resource Ecology, P.O. Box 510119, 01314 Dresden, Germany
[c.] Address here. Institute of Energy and Climate Research, IEK-6, Nuclear Waste Management and Reactor Safety, Forschungszentrum Jülich GmbH, Wilhelm-Johnen-Strasse, 52428 Jülich, Germany
[d.] JARA High-Performance Computing, Schinkelstraße 2, 52062 Aachen, Germany
[e.] Molecular and Condensed Matter Physics, Department of Physics and Astronomy, Uppsala University, SE-751 20 Uppsala, Sweden
[f.] Belgian Nuclear Research Centre (SCK·CEN), Institute for Nuclear Materials Science, Boeretang 200, B-2400 Mol, Belgium
[g.] Department of Applied Physics, Aalto University, P.O. Box 14100, FI-00076 Aalto, Finland

* These authors contributed equally.

**Valence band electronic structure of mixed uranium oxides ($UO_2$, $U_4O_9$, $U_3O_7$, $U_3O_8$, β-$UO_3$) has been studied by the resonant inelastic X-ray scattering (RIXS) technique at the U $M_5$ edge and by computational methods. We show here that the RIXS technique and recorded U 5f - O 2p charge transfer excitations can be used to proof the validity of theoretical approximations.**

Structural, electronic and chemical properties of uranium (U) oxides vary strongly upon a transformation from the fluorite-type $UO_2$ structures to the layered structure of the higher U oxides ($U_3O_8$ and above)[1–13]. The mechanism of the expansion of the fluorite structure is reasonably straightforward[14–16], however the role of oxygen (O) atoms in these structural changes remains less clear. We performed the state-of-art valence band RIXS experiment at the U $M_5$ edge for a number of binary U oxides – $UO_2$, $U_4O_9$, $U_3O_7$, $U_3O_8$, β-$UO_3$ – in order to clearly identify the mechanism causing the electronic structure modification upon oxidation of $UO_2$.

Valence band RIXS data at the U $M_5$ edge (~3550 eV) have been previously reported for $UO_2$, $UO_3$, $UF_4$, $UO_2(NO_3)_2 \cdot 6(H_2O)$ and several U intermetallic systems[17–20], and have been proved to be sensitive to the structural environment of U atom and its ligands. Actually, the valence band RIXS data include the elastic and inelastic scattering profiles with an energy resolution of ~1eV and provide information on the energy difference between the valence band states and the unoccupied U 5f states. Fig. 1 shows the valence band RIXS spectra of $UO_2$, $U_4O_9$, $U_3O_7$, $U_3O_8$ and β-$UO_3$, recorded at the Beamline ID26 of The European Synchrotron (ESRF)[21], (see ESI). The lowest energy feature at ~20 eV is attributed to the U $6p_{3/2}$-$3d_{5/2}$ transitions[17,18]. The process involves first the excitation of an electron from the U $3d_{5/2}$ core level (at the U $M_5$ edge) to the unoccupied U 5f state and then the U $3d_{5/2}$ core hole is filled by an electron from the occupied U $6p_{3/2}$ state. The inelastic scattering profile at energies transfer ~5-10 eV, reported in Fig.1 has been attributed to charge transfer process from the occupied O 2p states to the unoccupied U 5f states[18–20]. The energy separation between elastic and inelastic scattering contributions to the spectra depends of the energy difference between the occupied O 2p states and unoccupied U 5f states.

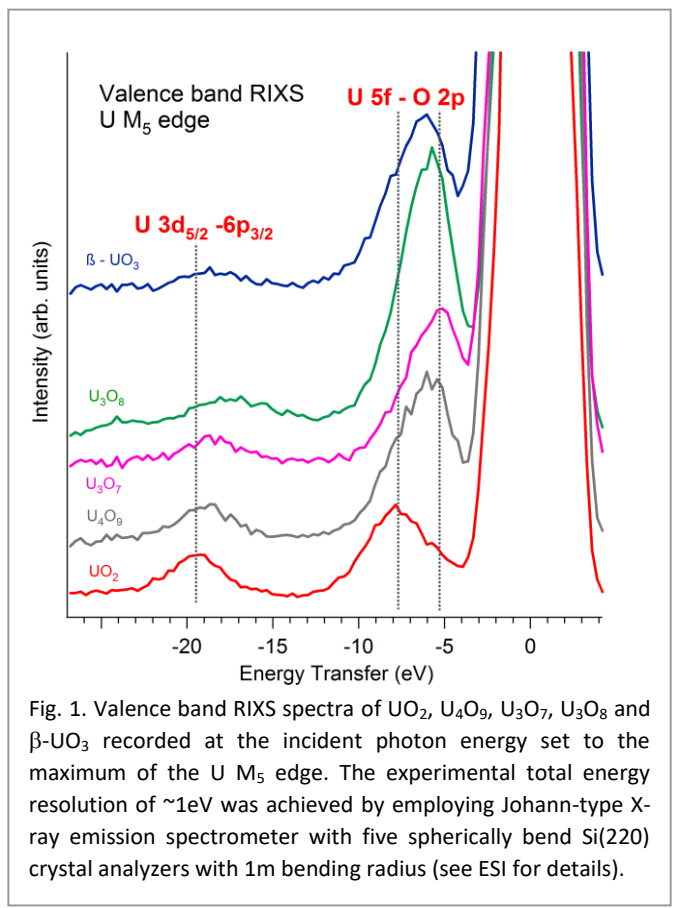

Fig. 1. Valence band RIXS spectra of $UO_2$, $U_4O_9$, $U_3O_7$, $U_3O_8$ and β-$UO_3$ recorded at the incident photon energy set to the maximum of the U $M_5$ edge. The experimental total energy resolution of ~1eV was achieved by employing Johann-type X-ray emission spectrometer with five spherically bend Si(220) crystal analyzers with 1m bending radius (see ESI for details).

Our recent studies of the evolution of the U chemical state in a series of U oxides confirms a changeover of the oxidation states U(IV) - U(V)- U(VI) through the charge compensation mechanisms[5,19,22]. The established formal oxidation states for U in mixed U oxides are included in brackets: $UO_2$(IV), $U_4O_9$(IV-V), $U_3O_7$(IV-V), $U_3O_8$(V-VI), β-$UO_3$(VI). Moreover the exact quantitative analysis has been performed and showed the presence of 50% and 50% of U(IV) and U(V), respectively, in $U_4O_9$; 33% and 67% of U(IV) and U(V) in $U_3O_7$ and 67% and 33% of U(V) and U(VI) in $U_3O_8$ [5].

Based on these findings the process of the electron transfer from the O 2p orbitals to the unfiled U 5f shell should show the constant increase of the energy separation between the elastic and inelastic scattering contributions in the spectra through the series of these mixed valence oxides. However, measured valence band RIXS data for the U oxides (Fig.1) show that the mechanism of the electronic structure modification during the transformation of $UO_2$ into the mixed oxides is however more complicated. Charge transfer also takes place between U sites and additionally incorporated $O^{2-}$ ions in binary oxides. As a result, the modification of the U-ligand bonding induces a change in the U oxidation state. In addition to that, the position and distribution of valence band states near the Fermi level changes significantly on a scale of several eV.

To gain better understanding and to clarify the mechanism of charge transfer excitations and electronic structure modifications we performed three types of calculations with methods of computational quantum chemistry, which have been used to evaluate the properties of U oxides materials previously [23–30].

Computation of mixed U oxides is a challenging task because of a strongly correlated and localized character of the *f* electrons[23,30]. The commonly used density functional (DFT) methods often fail even on the qualitative level, for instance predicting a metallic state for $UO_2$[23], which is in reality a Mott insulator with a wide band gap of 2.1 eV[31]. To correct for this, the DFT+*U* method is often used when an on-site Coulomb interaction is modelled by an additional term in the Hamiltonian, whose strength depends on the Hubbard *U* parameter [23,32,33]. This parameter in calculations for uranium oxides is usually taken as *U*=4.5 eV (with additional *J* parameter of 0.54eV) or an effective parameter, $U_{eff}$ =U-J, is applied[30,33]. These values come from the measurements of the correlation energy performed on $UO_2$ [34,35]. This approximation is the most common approach in the calculations of the electronic structure of U systems [10,19,23,25,29,36,37].

Recently, Beridze and Kowalski[30,38] performed systematic tests of the performance of the DFT+*U* method with the Hubbard *U* parameter derived *ab initio* using the linear response method of Cococcioni and Gironcoli[33] for the calculations of actinide bearing molecular and solid compounds, including U oxides. They have shown that the Hubbard *U* parameter values strongly depend on the oxidation state of U, being largest for U(VI) (~3 eV) and smallest for U(IV) (~2 eV). In follow up studies a similar trend has been shown for other actinides[38]. Here we test the performance of this methodology for the computation of the electronic density of states (DOS) that are used for the construction of theoretical RIXS data.

In most of the DFT+*U* implementations the shape of the orbitals of interest (*f* orbitals in the case of U) have to be provided in order to estimate the occupation of these orbitals and compute the Hubbard energy correction term[29,33]. These orbitals are usually represented by the atomic orbitals computed for atoms or ions and thus not necessarily adequately represent the shape of the orbital in a solid. In recent studies the maximally localized Wannier functions have been applied in computation of electronic structure of solids[32,38–40]. In this contribution we will test this approach in order to check if the Wannier functions-based representation of *f* orbitals can result

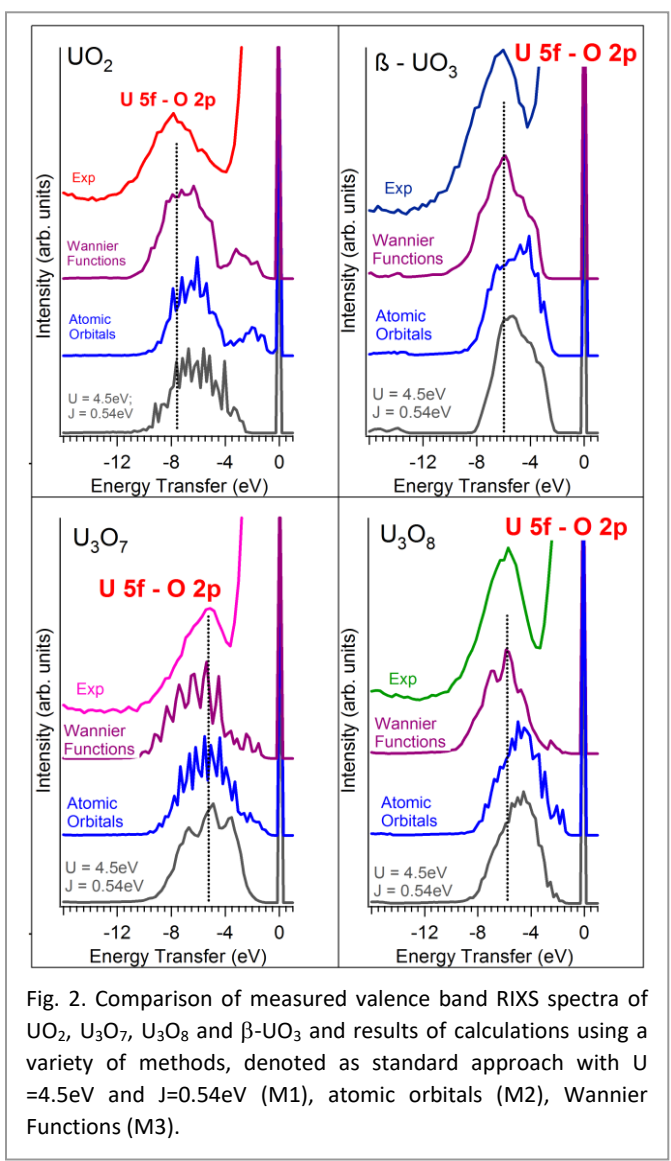

Fig. 2. Comparison of measured valence band RIXS spectra of $UO_2$, $U_3O_7$, $U_3O_8$ and β-$UO_3$ and results of calculations using a variety of methods, denoted as standard approach with U =4.5eV and J=0.54eV (M1), atomic orbitals (M2), Wannier Functions (M3).

in any significant improvement in the description of the DOS functions and RIXS data.

In order to reproduce the experimentally detected charge transfer excitations we made simplified calculations by inserting the calculated partial U 5f and O 2p DOSs into the Kramers-Heisenberg equation[36,41]. This approach provides straight forward information about the validity or accuracy of DOSs calculated using a variety of methods. It describes a correlation between occupied and unoccupied states under assumption that the hybridization between U 5f and O 2p states takes place. In that case, the energy difference between the maxima of the occupied O 2p DOS and unoccupied U 5f DOS will define the energy transfer values for the observed RIXS transitions.

The partial DOSs have been calculated by three approaches (see ESI). First we have applied the LDA+$U$ approach assuming the standard values used in calculations for U oxides (U=4.5eV and J=0.54eV)[23] (denoted as M1). In addition we computed the Hubbard $U$ parameters values using the linear response method of Cococcioni and Gironcoli [33]. Here, for the calculations of the Hubbard correction we represent $f$ orbitals for projection of occupations by the atomic orbitals (M2) and the Wannier functions (M3).

Computation of the electronic structure of U oxides can often lead to the convergence to a metastate instead of ground state[27]. In order to obtain the correct electronic structure of the considered oxides, for an initial electronic state we computed the expected charges of the different U atoms in the mixed U oxides using bond valence sum (BVS) method[42]. The BVS of U atoms for the considered oxides: $U_3O_7$ ($P4_2/n$)[13], $U_3O_8$ ($Amm2$), $UO_3$ ($P1211$), and $UO_2$ ($Fm$-$3m$) were calculated and analyzed applying the following formula:

$$V = \sum \exp[(R_j - d_j)/b]$$

Here the bond valence parameter $R_j$ and constant $b$ are taken from Ref.[43] V and $d_j$ are the corresponding valence and bond lengths for each phase. The BVS results are given in ESI, where the U average charge is in approximate agreement with our previous findings[5,22].

The Hubbard $U$ parameter values computed with the linear response for the considered U oxides are given in Table 1. In general, as in our previous studies[30], the value is smaller than 4.5 eV, but strong dependence on the oxidation state is observed. The largest value was obtained for $UO_3$ (U(VI)) and the smallest for $UO_2$ (U(IV)).

Fig. 2 shows a comparison of measured and calculated RIXS spectra of $UO_2$, $U_3O_7$, $U_3O_8$ and β-$UO_3$ at the maximum of the U $M_5$ edge, using three approximations (M1, M2 and M3). The elastic scattering contribution has been added[36] to the calculated RIXS spectra to facilitate a comparison with experimental data. The standard approach (M1) does not give an ideal match to the experimental RIXS data (inaccurate energy difference between elastic and inelastic theoretical RIXS profiles). The calculations with the Hubbard $U$ parameter derived $ab$ $initio$ (M2) are even more deviating from the experiment as a result of smaller Hubbard U values than the standard one (4.5 eV). The predicted band gaps (see Table 2) are also smaller than the measured ones[29,31]. On the other hand, the band gaps predicted by M1 method are in qualitative agreement with the measurements. It is important to note that U 5f - O 2p charge transfer excitations, recorded by RIXS in this case, show the energy difference between two electronic levels (empty U 5f and occupied O 2p) and do not directly related to the band gap values obtained by other experimental methods (like optical spectroscopy).

The issue of the DFT or DFT+$U$ predicted band gaps has been discussed previously[30]. One interesting aspect in the case of $UO_2$, which is often used as a model system, is that for the Mott insulator the band gap value should be well approximated by the value of the Hubbard $U$ parameter. The measured band gap for $UO_2$ is 2.1 eV, which is close to the Hubbard $U$ parameter value predicted by the linear response method (Table 1). The problem as outlined by Breidze and Kowalski[30] is that the atomic orbitals used to represent the $f$ orbitals in solids are not a good representation resulting in a significant and unrealistic fractional occupation of the unoccupied $f$ levels (up to 0.3 for $UO_2$). In order to remove this obstacle we applied the Wannier representations of the $f$ orbitals for the DFT+$U$ calculations, which resulted in more realistic, close to zero occupations of the unoccupied $f$ orbitals. The RIXS profiles resulting from the later calculations are also plotted in Fig. 2. These represent the best match to the measured RIXS profiles with good prediction of the position of the U 5f - O 2p charge transfer and an overall much better match to the observed shape of the U 5f – O 2p charge transfer excitations.

Table 1: The computed Hubbard U parameter values in eV

|  | U(IV) | U(V) | U(VI) |
|---|---|---|---|
| $UO_2$ | 1.7 | | |
| $U_3O_7$ | 2.1 | 2.1 | |
| $U_3O_8$ | | 2.0 | 2.2 |
| β-$UO_3$ | | | 2.5 |

Table 2: Predicted and measured band gaps (in eV) of the mixed U oxides.

|  | M1 (LDA + $U$) | M2 (Atomic) | M3 (Wannier) | Exp |
|---|---|---|---|---|
| $UO_2$ | 2.4 | 0.3 | 0.7 | 2.1 |
| $U_3O_7$ | 1.4 | 0.6 | 0.9 | 1.6 |
| $U_3O_8$ | 1.7 | 0.9 | 1.3 | 1.8 |
| β-$UO_3$ | 2.2 | 2.0 | 2.7 | 2.2 |

Table 3: The computed and measured energy difference between U 5f and O 2p states. The energies are given in eV

|  | M1 (LDA + $U$) | M2 (Atomic) | M3 (Wannier) | Exp |
|---|---|---|---|---|
| $UO_2$ | 6.5 | 6.3 | 7.0 | 7.5 |
| $U_3O_7$ | 5.8 | 5.6 | 6.5 | 5.5 |
| $U_3O_8$ | 5.2 | 5.3 | 6.1 | 6.0 |
| β-$UO_3$ | 5.2 | 5.5 | 6.5 | 6.2 |

We present here the electronic structure studies of several U mixed oxides. The RIXS spectra shown in Fig. 1 and Fig. 2 indicate the experimentally obtained energy between the occupied O 2p and unoccupied U 5f states. In Table 3 we report the difference between the average energy of these states as integrated from the computed DOS functions (see ESI) using the three theoretical methods. The best match is obtained with the method M3 (with Wannier functions). Here the experimentally observed trend is clearly reproduced with the largest differences for $UO_2$ and the smallest for $U_3O_8$. These differences result from the decrease of the energy of the O $p$ states (with respect to the Fermi level) with increasing the oxidation state (due to the stronger electrons binding) and associated decrease of the energy of the unoccupied 5f states. The later effect results from higher Hubbard *U* parameter values (strength of the on-site Coulomb repulsion) for higher oxidation states of U (Table1). These interesting results indicate that with such a state-of-art experimental method – valence band RIXS - one can improve the theoretical prediction of the electronic structure of actinide contained materials.

**Notes and references** P.M.K. and J.L. thank the JARA-HPC for giving time on the supercomputing resources awarded through JARA-HPC Partition. S.M.B. acknowledges support from the Swedish Research Council (research grant 2017-06465) and K.O.K acknowledges support from ERC (research grant 759696).